\documentclass{iau}

\usepackage{amsmath}
\usepackage{graphicx}
\usepackage{multirow}

\def\mf{meridional flow}
\def\dr{differential rotation}

\def\we{Waldmeier effect}
\def\bl{BL}

\newcommand{\Fig}[1]{Figure~\ref{#1}}

\begin{document}

\lefttitle{Bidya Binay Karak}
\righttitle{Recent Developments in the Babcock-Leighton Solar Dynamo Theory}

\jnlPage{1}{7}
\jnlDoiYr{2021}
\doival{10.1017/xxxxx}

\aopheadtitle{Proceedings IAU Symposium}
\editors{A. V. Getling \&  L. L. Kitchatinov, eds.}

\title{Recent Developments\\ in the Babcock-Leighton Solar Dynamo Theory}

\author{Bidya Binay Karak}
\affiliation{Department of Physics, Indian Institute of Technology (Banaras Hindu University), Varanasi 221005, India\\
email:\email{karak.phy@iitbhu.ac.in}}

\begin{abstract}
Babcock-Leighton process, in which the poloidal field is generated through the decay and dispersal of tilted bipolar magnetic regions (BMRs), is observed to be the major process behind the generating poloidal field in the Sun. Based on this process, the Babcock-Leighton dynamo models have been a promising tool for explaining various aspects of solar and stellar magnetic cycles. In recent years, in the toroidal to poloidal part of this dynamo loop, various nonlinear mechanisms, namely the flux loss through the magnetic buoyancy in the formation of BMRs, latitude quenching, tilt quenching, and inflows around BMRs, have been identified. While these nonlinearities tend to produce a stable magnetic cycle, the irregular properties of BMR, mainly the scatter around Joy's law tilt, make a considerable variation in the solar cycle, including grand minima and maxima. After reviewing recent developments in these topics, I end the presentation by discussing the recent progress in making the early prediction of the solar cycle.
\end{abstract}


\maketitle

\section{Introduction}
Many solar-type stars (spectral type: F8V--K2V and effective temperature: 5100--6000~K) show 
cyclic variations of their magnetic fields \citep{Baliu95,garg19}. The (large-scale) magnetic field of our Sun 
flips every 11 years, and the strength of this field in each cycle varies cycle-to-cycle
which is popularly measured by the number of sunspots \citep{Hat15, Biswas23}. 
While there are variations within 
the cycle, such as variable rise rate \citep[which is related to the cycle strength and is given by 
Waldmeier effect;][]{KC11} and Gnevyshev peaks \citep{KMB18}, there are variations beyond 11-year periodicity
such as the Gnevyshev–Ohl rule, grand minima/maxima and Gleissberg cycle \citep{Uso23}.

It is observed that the magnetic field of the Sun is produced through the dynamo process, which 
is a cyclic conversion between the toroidal and poloidal fields \citep{Cha20}. A strong toroidal magnetic field is produced 
from the poloidal component due to the shearing by the differential rotation in the solar convection zone (SCZ).
Due to magnetic buoyancy, this toroidal magnetic field gives rise to bipolar magnetic regions (BMRs) 
or loosely sunspots. The decay and dispersal of these BMRs give a poloidal magnetic field, which was first proposed by \citet{Ba61} and \citet{Le64} and is the heart of surface flux transport (SFT) models \citep{Kar14a}. 
While $\alpha$ effect---the lifting and twisting of toroidal field due to helical convection---is 
another obvious candidate for the generation of poloidal field, the Babcock--Leighton (\bl) process is observationally supported 
and is the major process for generating the poloidal field in the Sun \citep{CS23}. 
We note that in the \bl\ process, it is the tilt of the BMR which is crucial in generating the poloidal field in the Sun. 
\mf\ also helps in transporting the field from low to high latitudes (largely by dragging the trailing polarity flux of BMRs), 
producing the observed pole-ward migration of surface field \citep{Bau04, Mord22}.

In the basic BL dynamo framework, we expect that the poloidal to toroidal field conversion part is fairly understood due to the following facts. (i) The observed \dr\ in Sun is observed to vary only by a little amount (in the form of torsional oscillation). (ii) Polar field (or its proxy) is linearly correlation with the amplitude of the next cycle \citep{WS09, KO11, Muno13, Priy14, Kumar21, Kumar22}. (iii) Toroidal flux produced due to shearing of the observed polar flux through the differential rotation matches well with the observation \citep{CS15}. However, the toroidal to poloidal part of the model was less transparent as this process involves some nonlinearity and stochasticity. In recent years, we have made some progress in understanding this part, which I shall highlight in this proceedings.

\section{Nonlinearities in \bl\ process}
\begin{figure}[t]
\begin{minipage}[t]{0.55\textwidth}
  \centering
  \includegraphics[scale=.3]{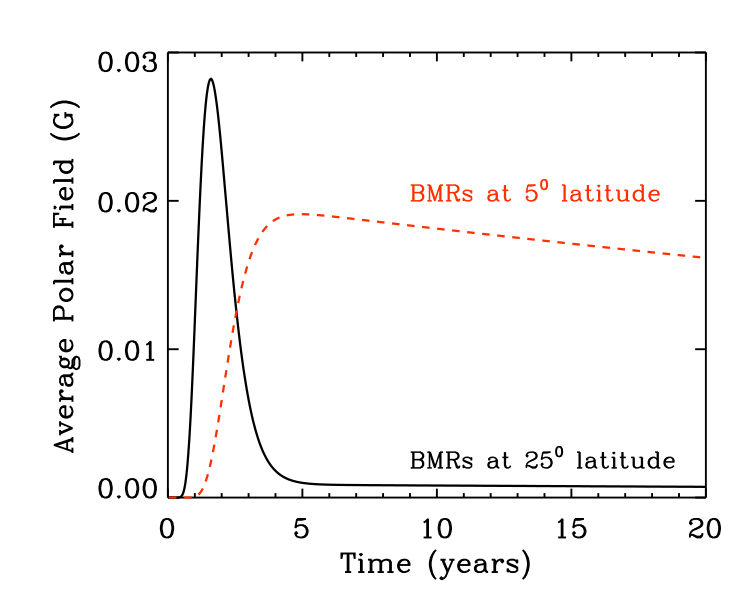}
  \end{minipage}%
  \begin{minipage}[t]{0.45\textwidth}
  \vspace{-4.2cm}
  \caption{The dashed/red line shows the polar field produced from the decay of two BMRs deposited symmetrically at $5^\circ$  north and south. Solid/black line represents the same from the same two BMRs but deposited at $5^\circ$ north and south,  and the tilts are assigned according to Joy's law. This plot shows the evidence of latitude quenching---BMRs appearing at high latitude generates less polar field.}
  \end{minipage}
  \label{fig:lat_quench}
\end{figure}

There are several nonlinear processes involved in the \bl\ mechanism for the generation of the poloidal field in the Sun. To explain these, let us begin by observing the butterfly diagram \citep[Figure 9][]{Hat15}. We find that the strong cycles begin BMRs/sunspots at high latitudes, and thus, the average latitudes of BMR are
high for strong cycles and vice versa \citep{MKB17}. On the other hand, we find that if a BMR appears at a high latitude, it gives less polar field than the BMR at a low latitude; see \Fig{fig:lat_quench}. Hence, suppose due to fluctuations, if the Sun's magnetic field in a cycle is trying to grow, it will produce BMRs at high latitudes (as observed in the Sun). These high-latitude BMRs will give less polar field. The less polar field will consequently create a less toroidal field for the next cycle, and the indefinite growth of the magnetic field will be halted. This mechanism, so-called latitude quenching as proposed by \citet{Petrovay20} and \citet{J20} is shown to be a potential mechanism for stabilizing the magnetic field in the solar dynamo as demonstrated by \citet{Kar20} in 3D \bl\ dynamo model \citep{MD14, MT16, KM17}. 

Strong cycles produce BMRs at higher latitudes, related to a broader feature of the sunspot butterfly diagram, first highlighted by \citet{W55} and recently by \citet{CS16}. 
They show that a strong cycle begins BMRs at higher latitudes; its activity level (sunspot number) rises rapidly (in the activity vs central latitude of the latitudinal distribution of sunspots), and it begins to decline already when the sunspot activity belt is at high latitudes. This feature was explained by \citet{BKC22} by including a  flux loss due to magnetic buoyancy through the formation of BMRs in their dynamo model \citep[following the idea of][including the magnetic buoyancy in axisymmetric dynamo model]{NC00}. They showed that strong cycles produce BMRs at a high rate, and thus, they lose their toroidal flux rapidly. This rapid loss of toroidal flux causes the cycle to decline early when the toroidal belt is at high latitudes.

Another nonlinearity in the \bl\ solar dynamo is the reduction of tilt with the magnetic field, so-called tilt quenching. 
In the thin flux tube theory of BMR formation, the tilt of a BMR is produced due to the torque induced by the Coriolis force on the diverging flow produced from the apex of the rising flux tube \citep{DC93}. Hence, we expect that if the BMR forming flux tube has a strong magnetic field, it will rise quickly, and the Coriolis force will take less time to develop tilt. Thus, we expect the BMR tilt to decrease with the increase of the magnetic field in BMR. Based on this idea, kinematic \bl\ 
dynamo models use a magnetic field-dependent quenching in the tilt \citep{LC17, KM17, KM18}. However, the observational signature of this tilt quenching is fragile due to limited magnetogram data of strong cycles \citep{Jha20}; also see \citet{Jiao21} for the cycle averaged tilt from white-light data.

The final nonlinearity that we shall discuss here is the inflows around BMRs. Observations show converging flows around the BMRs \citep{Gizon01,Gonz08}. These inflows aggregately generate mean flows around the activity belt \citep{Jiang10, CS12}.
Due to these flows, the cross-equatorial cancellations of the BMRs are reduced, and the effectivity of the \bl\  process is suppressed. In a strong cycle, this effect is stronger and thus leads to a stabilizing effect in the dynamo \citep{MC17, Nagy20, Kinfe23}. 

Besides the above nonlinearities in the toroidal to poloidal part of the dynamo model, the poloidal to toroidal part and the turbulent transport coefficients are also nonlinear. However, due to the fact that the \dr\ is observed to vary only weakly with the solar cycle, and there is a strong correlation between the polar field and the next cycle amplitude, we expect that the poloidal to toroidal part is weakly nonlinear. We refer readers to Section 5 of \citet{Kar23} for all possible nonlinearities in the solar dynamo.

 \begin{figure}[t]
  \centering
  \includegraphics[scale=.25]{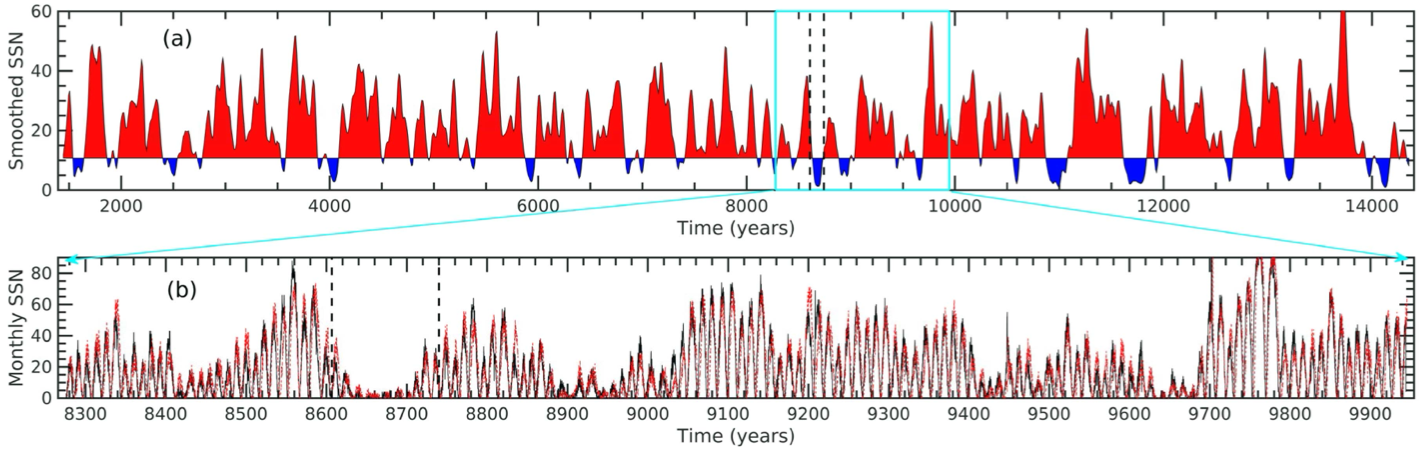}
  \caption{(a) Smoothed (10-year binned and smoothed) BMR number from 3D dynamo model of \citet{KM18}. Blue regions show the locations of grand minima. (b) shows the zoomed-in view of some cycles. A Maunder-like extended grand minimum is highlighted by dashed lines. Note that the model successfully recovers from grand minima without needing any additional source for the poloidal field.}
  \label{fig:grandmin}
\end{figure}

\section{Stochastic effects in \bl\ process}
The nonlinearities mentioned above in the \bl\ dynamo model have the tendency to stabilize the magnetic cycle rather than produce irregularity. However, the solar cycle is irregular. Hence, can the \bl\ dynamo models produce 
the observed modulations in the solar cycle, including grand minima and maxima? The answer is again hidden in the \bl\ 
process. As mentioned in the introduction, the tilt of BMR plays a vital role in generating a poloidal field in the sun. While this 
tilt follows Joy's law in a statistical sense, there is a huge scatter around it which can be approximated by a Gaussian of sigma about $19^\circ$ \citep{SK12,Jha20,Anu23}. Not only the BMR tilt but the BMR eruption rate and flux content also give some variations to the magnetic cycle \citep{KM17, Kumar23}. By including the scatter around Joy's law tilt and variation in the BMR flux and eruption rates \citet{KM17} showed modulation in the solar cycle, including grand minima and maxima episodes (\Fig{fig:grandmin}); also see \citet{LC17} for a similar work. Several previous studies were also performed to model the grand minima using axisymmetric dynamo models where the BMR scatter was captured by including noise in the \bl\ source term \citep[e.g.,][]{CK09, CK12, OK13, KC13, Pas14, Nagy17}. Double peaks (Gnevyshev peaks) were also modelled using a similar idea \citep{KMB18}. Recently, \citet{Elena23, BKK23} also showed that the scatter around BMRs, particularly the anti-Hale BMRs, produces considerable variation in the polar field strength and the reversal times. 

\begin{figure}[t]
  \centering
  \includegraphics[scale=.235]{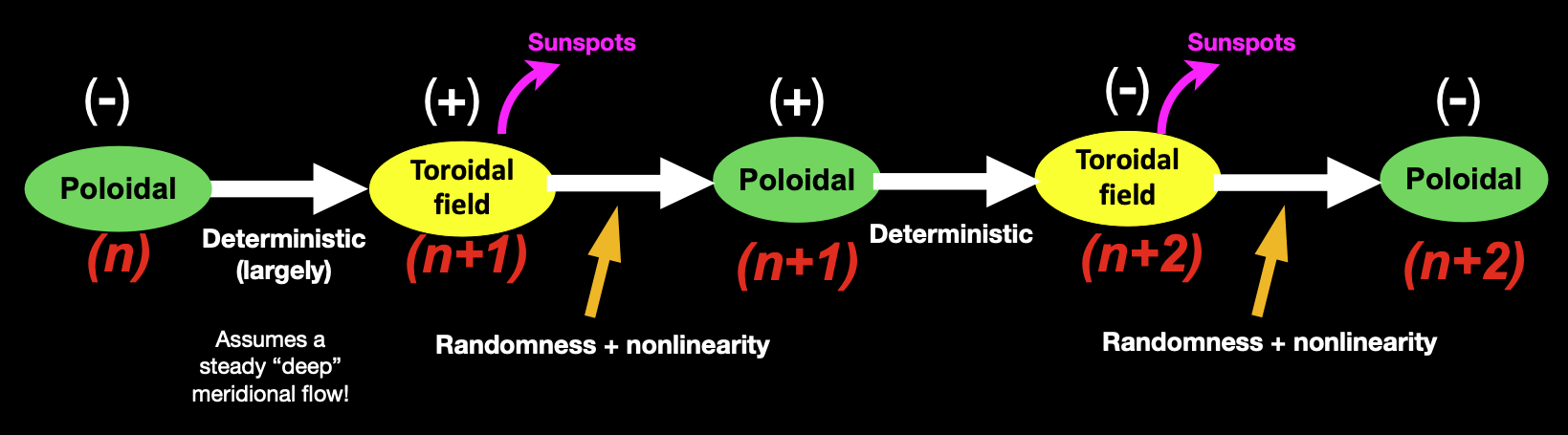}
  \caption{Dynamo chain under the \bl\ framework.}
  \label{fig:chain}
\end{figure}

\begin{figure}[t]
\begin{minipage}[t]{0.77\textwidth}
  \centering
  \includegraphics[scale=.45]{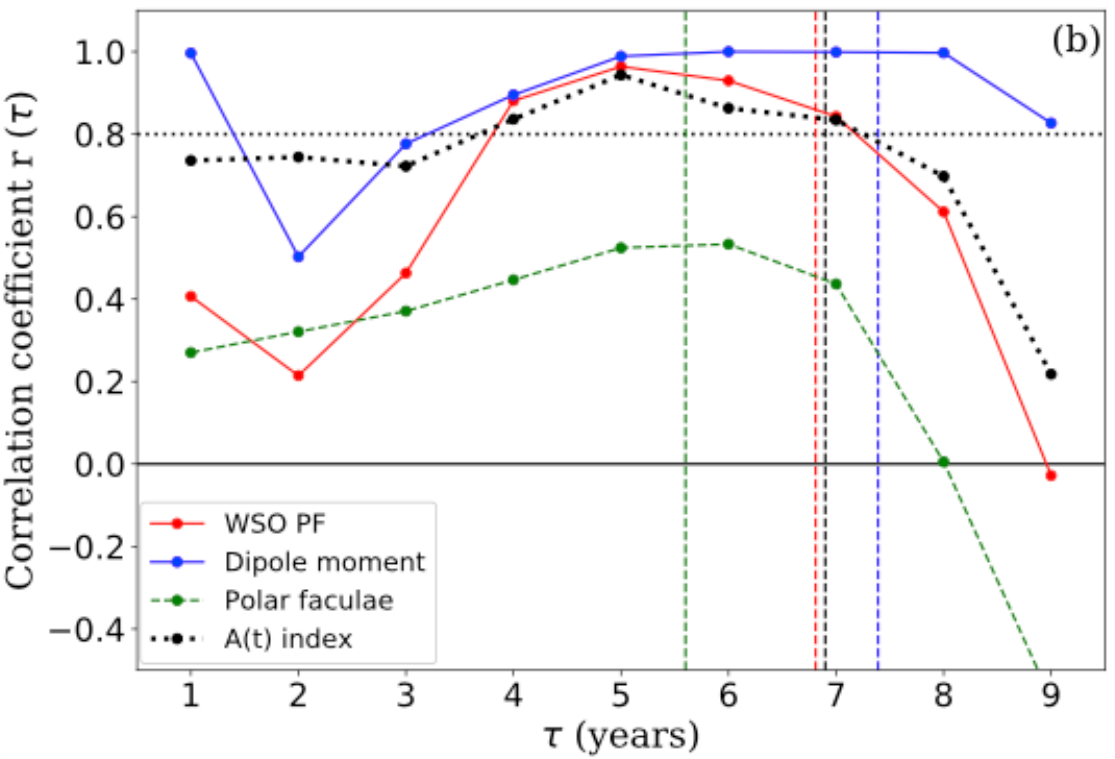}
  \end{minipage}%
\begin{minipage}[t]{0.23\textwidth}
\vspace{-6.0cm}
  \caption{Linear correlation coefficient between the amplitude of the next sunspot cycle and the polar field or its proxy at different times $\tau$ measured from the times of reversals. Dashed lines show the time difference between the corresponding time of polar field reversal and the cycle minimum. Figure reproduced from \citet{Kumar21}.}
  \end{minipage}
  \label{kumar_lagcorr}
\end{figure}

Another potential source for the irregularity in the solar cycle is the variation in the meridional circulation. Surface observations
show a noticeable temporal variation of it over the last few solar cycles \citep[e.g.,][]{Gonz08}. However, the variation in the deep meridional circulation is uncertain due to the difficulties in the measurement. \citet{Kar10} have shown that a weak \mf\ makes a cycle long and magnetic field weak (due to getting more time to diffuse the field) in the high diffusivity dynamo model. He showed 
that a large part of the amplitude is also matched by modelling the cycle period by varying the \mf\ speed. Later, \citet{KC11} showed that a variable meridional circulation helps in modelling the classical \we. A large reduction in the meridional flow also allows the model to push into a grand minimum phase \citep{Kar10, KC12, KC13}. SFT models also show that variation
in the \mf\ cause variation in the polar field \citep{UH14b, BKK23}.

\section{Early prediction of solar cycle}
\begin{figure}[t]
\begin{minipage}[t]{0.77\textwidth}
  \centering
  \includegraphics[scale=.23]{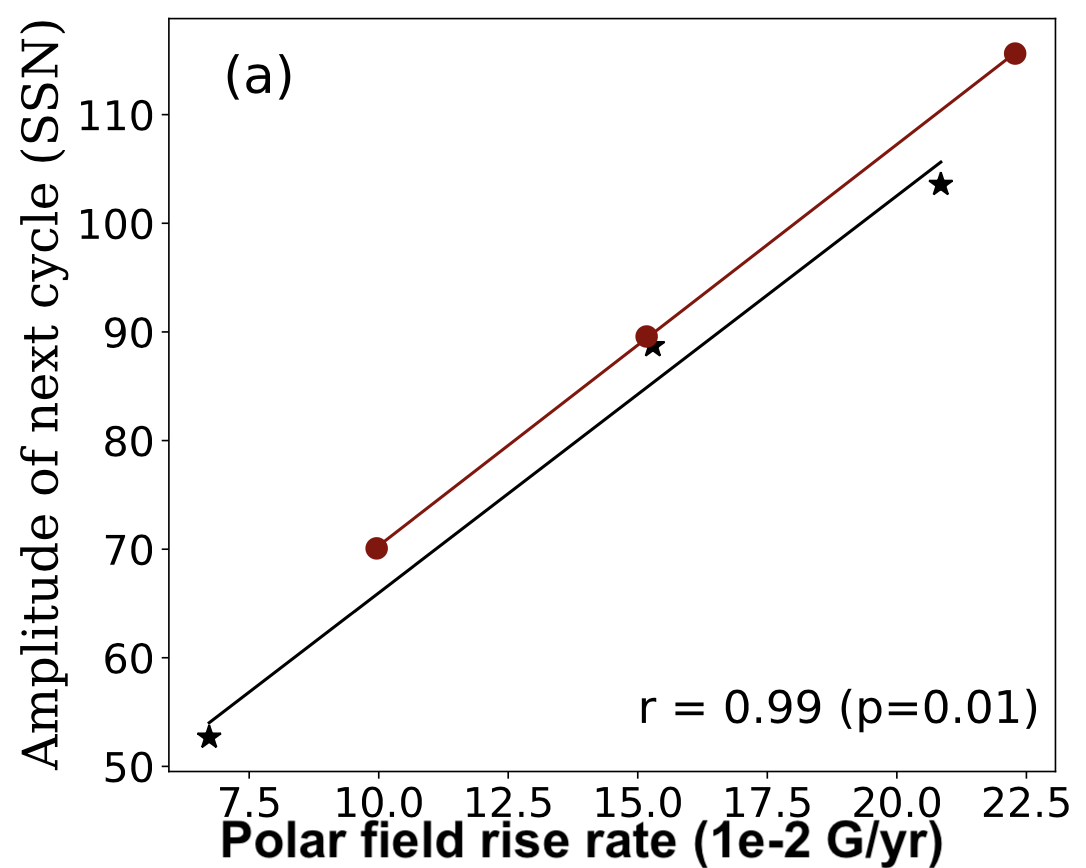}
  \includegraphics[scale=.33]{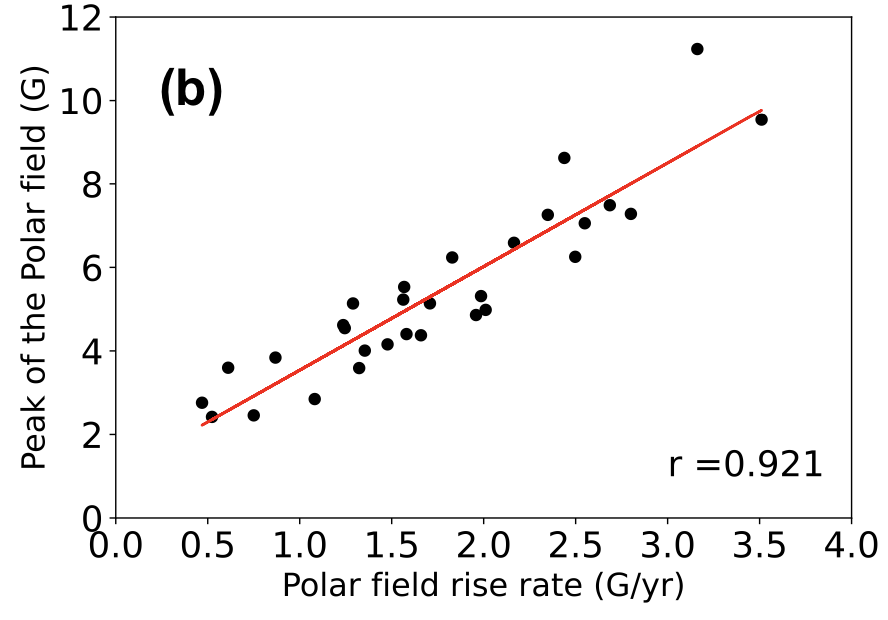}
  \end{minipage}%
\begin{minipage}[t]{0.23\textwidth}
\vspace{-3.8cm}
  \caption{(a) Correlation between the observed rise rate of the polar field with the amplitude of the next sunspot cycle \citep{Kumar22}. (b) Same as (a) but obtained from an SFT model \citep{BKK23}.}
  \end{minipage}
  \label{fig:riseratecorr}
\end{figure}

\begin{figure}[t]
\begin{minipage}[t]{0.79\textwidth}
  \includegraphics[scale=.25]{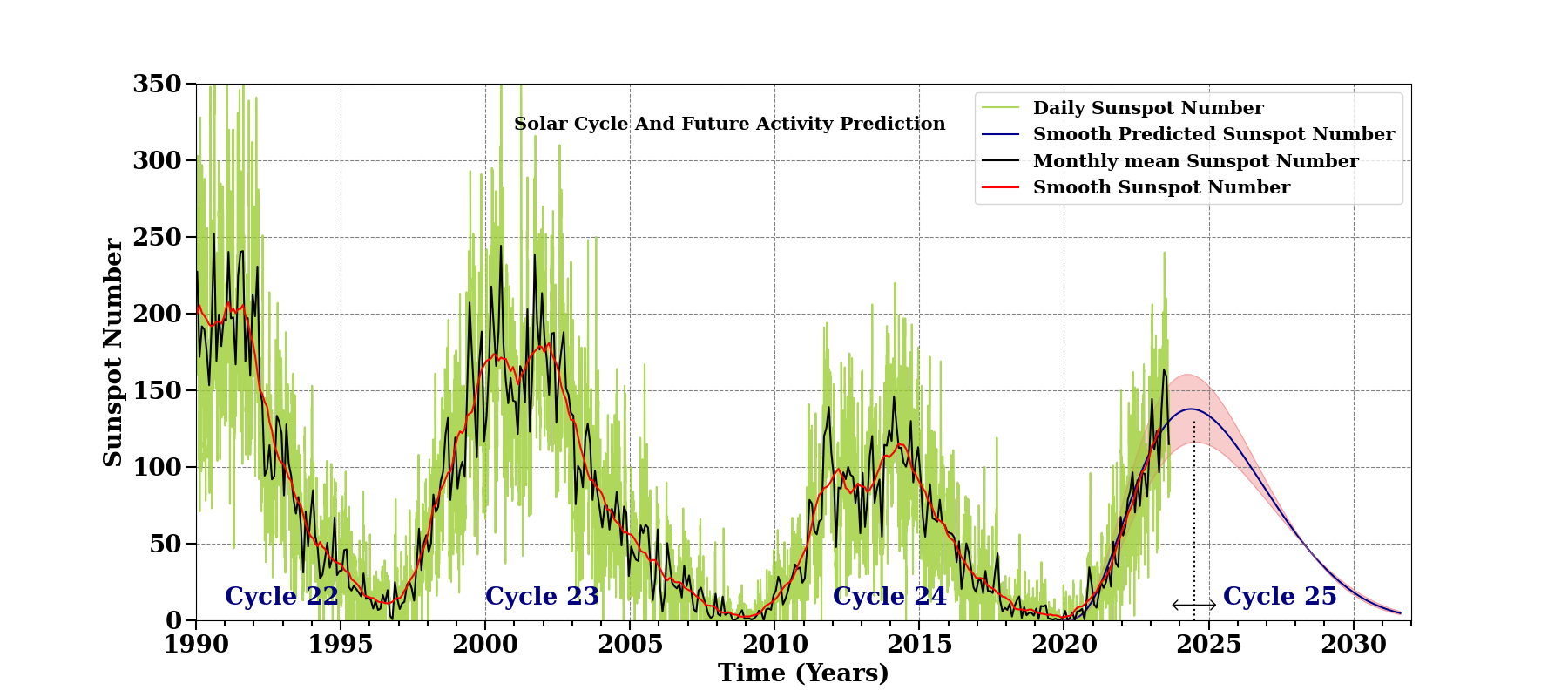}
    \end{minipage}%
\begin{minipage}[t]{0.21\textwidth}
\vspace{-4.7cm}
  \caption{Prediction of the Cycle 25 using the rise rate of the polar field during three years after reversal \citep{Kumar22}. The red-shaded region shows the prediction window, and the horizontal arrow shows the range of expected peak time.}
    \end{minipage}
  \label{fig:prediction}
\end{figure}

We have seen that the solar dynamo has a stochastic component, and it is nonlinear. So the question is, can we then make a prediction of the future cycle? To answer this question, let us recall the dynamo chain of the solar cycle as shown in \Fig{fig:chain}. We observe that the generation of the poloidal field involves some randomness, and thus, at the end of every cycle, the strength of the poloidal field varies from cycle to cycle. However, the poloidal to toroidal part of the dynamo is deterministic (as the differential rotation in the Sun does not change much with time). The toroidal magnetic field gives sunspots for the next cycle. In fact, this is the basis for producing the observed correlation between the poloidal field (or its proxy) and the next sunspot cycle. Hence, if we can measure
the poloidal magnetic field, we can predict the amplitude of the next cycle. Indeed, a part of the poloidal field, namely the radial field on the solar surface, is observed, and it becomes maximum near the solar minimum. Prediction works use the polar field at the solar minimum to predict the amplitude of the next cycle either by using the linear correlation between these two or by feeding the polar field in the dynamo model \citep{CCJ07, JCC07}. As the polar field peaks around the cycle minimum, most of the prediction methods use the polar field at the cycle minimum to predict the amplitude of the next cycle \citep{Bhowmik23}. However, the question is, `Do we really need to wait till the solar minimum or a reasonable prediction can be made sometime earlier?'  \citet{Kumar21} have shown that the correlation between the amplitude of the next cycle and the polar field just after four years becomes strong; see \Fig{kumar_lagcorr}. This suggests that a reasonable prediction of the amplitude of the 
next solar cycle can be made just after four years of the reversal, which is about 2--3 years before the solar minimum (the usual landmark of solar cycle prediction). Later \citet{Kumar22} have shown that instead of computing the polar field at a fixed time, if
we compute the rise rate (slope) of the polar field within the first three years of the polar field after the reversal, then a better and even early prediction can be made.
 They further showed that the rise rate of the polar field determines its eventual peak (\Fig{fig:riseratecorr}a)---it is like 
the \we\ in the polar field. While their initial inference was based on the limited observed data of the polar field, later \citet{BKK23} demonstrated the robustness of this feature using SFT model by feeding the synthetic solar cycle with anti-Hale and non-Joy BMRs; see \Fig{fig:riseratecorr}(b). Following this idea, \citet{Kumar22} forecasted the solar cycle 25 with predicted peak sunspot number: $137\pm23$;  see  \Fig{fig:prediction}.

\begin{figure}[t]
  \includegraphics[scale=.48]{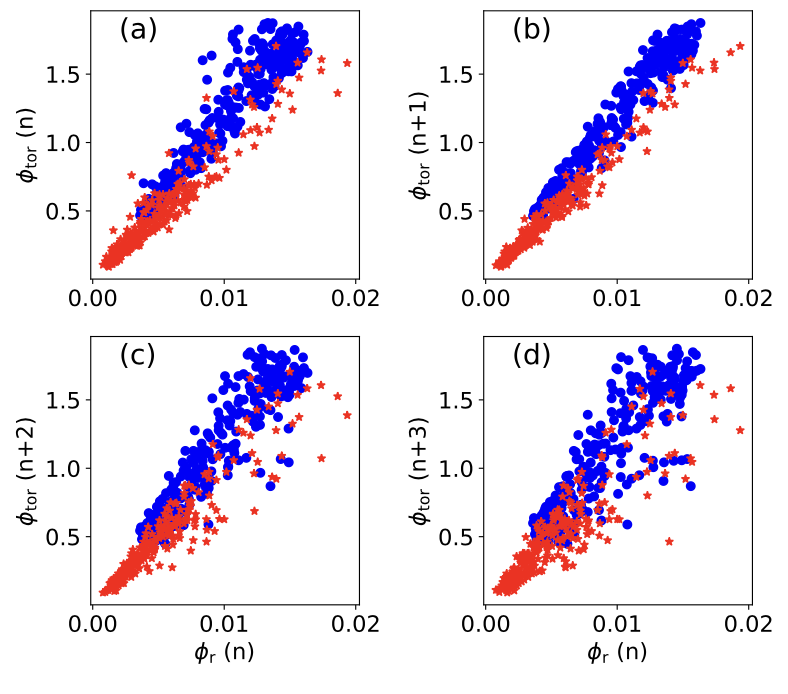}
    \includegraphics[scale=.48]{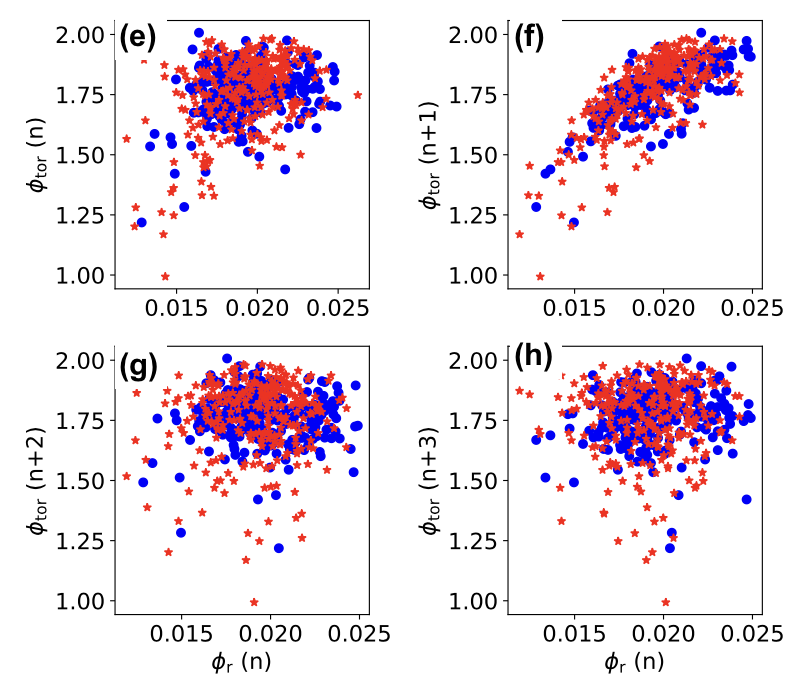}
  \caption{(a--d): Correlations between the polar field of cycle $n$ with the toroidal flux of cycle $n$, $n+1$, $n+2$, and $n+3$, respectively, when the dynamo operates near-critical regime (two times the critical dynamo number). (e--f) Same as (a--d) but when the model operates at a highly supercritical regime (four times critical); for details, see \citet{Kumar21b} from where this figure is adopted.}
  \label{fig:polcorr}
\end{figure}

Finally, the question is, if we can predict the amplitude of the next cycle, can we extend this further to multiple cycles? 
The simple answer is no. As seen from \Fig{fig:chain}, the toroidal to poloidal part of the solar dynamo diminishes the memory
of the polar field due to randomness and nonlinearity involved in this step. \citet{Kumar21b} have shown that only when the 
dynamo operates near the critical transition, the nonlinearity is weak, the dynamo growth rate is weak, and the memory of the polar field can be propagated
to multiple following cycles. With the increase of supercriticality of the solar dynamo, the memory of the polar field reduces, and in a highly supercritical regime, the memory is completely washed out; see \Fig{fig:polcorr} for an illustration. In conclusion, a reliable prediction of the solar cycle amplitude beyond one cycle is impossible. However, a hint of the amplitude of cycle $n+2$ can be obtained from the polar field of cycle $n$ as some studies suggest the solar dynamo is operating in weakly supercritical region 
\citep{KKB15, KN17, CS17, Vindya21, tripathi21, albert21, V23}.

\section{Concluding remarks and future outlook}
In recent years, we have had a tremendous development in the \bl\ dynamo theory for modelling the various aspects of the solar magnetic cycle. It has been shown that the toroidal to poloidal field part of the \bl\ dynamo model includes some nonlinearities, which at least include the flux loss due to magnetic buoyancy during the formation of BMR, latitude quenching, tilt quenching, and magnetic field dependent inflows around BMRs. These nonlinearities tend to stabilize the dynamo rather than produce a considerable variation in the magnetic cycle. Stochastic properties in the BMRs, namely, the scatter in the BMR tilt around Joy's law, can produce large variations in the solar cycle, including grand minima and maxima. Despite these nonlinearities and stochasticity in the solar dynamo, a prediction of the amplitude of the next solar cycle can be made using the polar field at the solar minimum. We have shown that we indeed do not have to wait till the minimum of the cycle; we can make a prediction of the next cycle's strength just by measuring the rise rate of the polar field of the first three years' data after its reversal. This allows us to make an early prediction of the solar cycle---about three years before the solar minimum, the usual landmark of the forecast. Prediction of multiple cycles is not possible because the memory of the polar field degrades in the poloidal to toroidal part of the dynamo loop. How rapidly the memory degrades beyond one cycle is determined by the supercriticality of the dynamo.

We end the discussion by commenting on the operation of \bl\ dynamo during Maunder-like grand minima. 
It has been suspected that the generation of the poloidal field through \bl\ process during these episodes is not possible due to
lack of sunspots. The obvious candidate for this is the classical $\alpha$ effect \citep{Pa55}. Some studies have included $\alpha$ effect in the \bl\ dynamo models to recover the models from these quiescent episodes \citep{KC13, Ha14, olc19}. However, 
recent observations show that the Maunder minimum was not as deep as it was thought earlier \citep[e.g.,][]{Uso15, ZP16}. Also, the small sunspots and the BMRs, which do not appear as sunspots in today's telescope \citep{Jha20}, were not observed during Maunder minimum. 
These BMRs also show a systematic tilt and Joy's law \citep{Jha20}, which can generate some poloidal field during Maunder-like grand minimum as demonstrated by \citet{KM18}. The downward turbulent pumping helps in reducing the diffusion of the poloidal 
field \citep{KC16}, and thus, even a weak poloidal field produced through the decay of a few BMRs during the grand minimum can successfully recover the Sun to normal phase. Another comment about the operation of \bl\ dynamo in other solar-type stars. Possibly, the BMRs are also produced in other stars, and they are tilted. With the increase of rotation rate, both the tilt and latitude of the emergence of BMR increase \citep{SS92}. However, the tilt cannot increase beyond $90^\circ$. This can cause saturation of magnetic field in rapidly rotating stars \citep{KO16}. The increase in the latitude of emergence can cause a decrease in the magnetic field \citep[latitude quenching;][]{J20, Kar20}. \citet{KKC14, Hazra19, KTV20, Vindya21, V23} have utilized \bl\  dynamo models to explain some aspects of the stellar magnetic cycles, including the saturation of the magnetic field in rapidly rotating stars and the variation of the number of grand minima and cycle variability with the rotation rates of stars. We hope future observations will help to validate the \bl\ dynamo theory in other stars.

\section{Acknowledgements}
Author thanks the International Astronomical Union for the travel grant to attend the IAU Symposium 365 in Yerevan, Armenia. The partial support from the Ramanujan Fellowship (project no SB/S2/RJN-017/2018) is gratefully acknowledged.

\bibliographystyle{iaulike}
\bibliography{iaukarak}
\end{document}